\documentclass{nature2}
\usepackage{graphicx}
\usepackage[usenames]{color}
\usepackage{xcolor}
\usepackage[export]{adjustbox}
\usepackage{amsmath}
\usepackage{amssymb}
\usepackage{chemformula,array}
\usepackage[symbol]{footmisc}

\title{Responsive Disorder in a Metal--Organic Framework\\ Enables Solid-State Reservoir Computing}

\author{Guy Greenbaum,$^1$ Will R. Branford,$^{2,3}$ and Andrew L. Goodwin$^{1\ast}$}

\begin{document}

\maketitle

\begin{affiliations}
	\item Inorganic Chemistry Laboratory, Department of Chemistry, University of Oxford, Oxford, UK
    \item Blackett Laboratory, Imperial College London, London, UK
    \item London Centre for Nanotechnology, Imperial College London, London, UK
\end{affiliations}

\begin{abstract}
Complex systems with nonlinear response mechanisms can be applied as reservoir computers for energy-efficient machine learning tasks. Historically explored at the macro- and meso-scale, physical reservoir computing has recently been extended to the atomic scale \emph{via} chemical mixtures with strong and dynamic heterogeneity. Here we explore the possibility that configurational degeneracy within disordered materials might form the basis for solid-state atomic-scale reservoirs. Our proof-of-concept uses the disordered metal--organic framework DUT-8, which undergoes a series of disorder--disorder transitions on exposure to different guest species. We show that variations in X-ray diffuse scattering associated with these transitions function as suitable readouts for machine learning applications. A combination of nonlinearity and memory effects in the DUT-8 response allows the system to carry out both classification and time-series machine learning tasks with accuracies comparable to those of mesoscale physical reservoir computers. Our results suggest a new avenue for exploiting correlated disorder in solid phases whenever the nature of that disorder can be modulated through external perturbations---a phenomenon we term `responsive disorder'.
\end{abstract}

\section*{Introduction}

Structural disorder can impart complexity (in the information-theoretic sense) whenever it is not random.\cite{Weaver_1948,Crutchfield_2011} Such non-random, or `correlated', disorder\cite{Keen_2015} has long been known across materials classes as diverse as solid electrolytes,\cite{Nield_1993} frustrated magnets,\cite{Fennell_2009} and supramolecular assemblies,\cite{Blunt_2008} but it is becoming increasingly apparent that the reticular approach of metal--organic framework (MOF) chemistry offers especially attractive opportunities for its rational design.\cite{Meekel_2021} In the nickel-containing MOF DUT-8,\cite{Klein_2010} for example, a combination of linker-orientation degrees of freedom and framework-imposed topological constraints leads to a highly degenerate and strongly correlated manifold of accessible disordered structures,\cite{Reynolds_2021} collectively representing a configurational landscape described in complexity theory as the `six-vertex' model.\cite{Lieb_1967} Many other complex architectures are possible, including examples related to so-called Truchet tilings\cite{Meekel_2023} and well-known statistical mechanical models as the Ising triangular antiferromagnet.\cite{Griffin_2025}

A particular hallmark of complex systems is their capacity for non-linear responses to external stimuli,\cite{Crutchfield_1994} and it is precisely this aspect that is exploited in neuromorphic computing.\cite{Mandic_2001} We are particularly interested in the development of so-called reservoir computing---a neuromorphic-inspired approach developed to address the enormous energy cost of traditional machine learning models.\cite{Jaeger_2004} Reservoir computers replace fully trainable neural networks with a fixed, dynamic `reservoir', using only a single trainable output layer to map reservoir readouts to desired outputs. Reservoirs that satisfy key criteria of nonlinearity and memory---\emph{i.e.}\ history-dependent response---retain many of the predictive capabilities of conventional neural networks, despite operating at a fraction of the training cost.\cite{Lukosevicius_2009} Because the connections in the reservoir are fixed, this architecture is compatible with physical implementations,\cite{Tanaka_2019} with examples including electrical circuits,\cite{Soriano_2015} tensegrity-based robots,\cite{Caluwaerts_2014} and even artificial limbs.\cite{Nakajima_2015}

\begin{figure}
    \centering
    \includegraphics{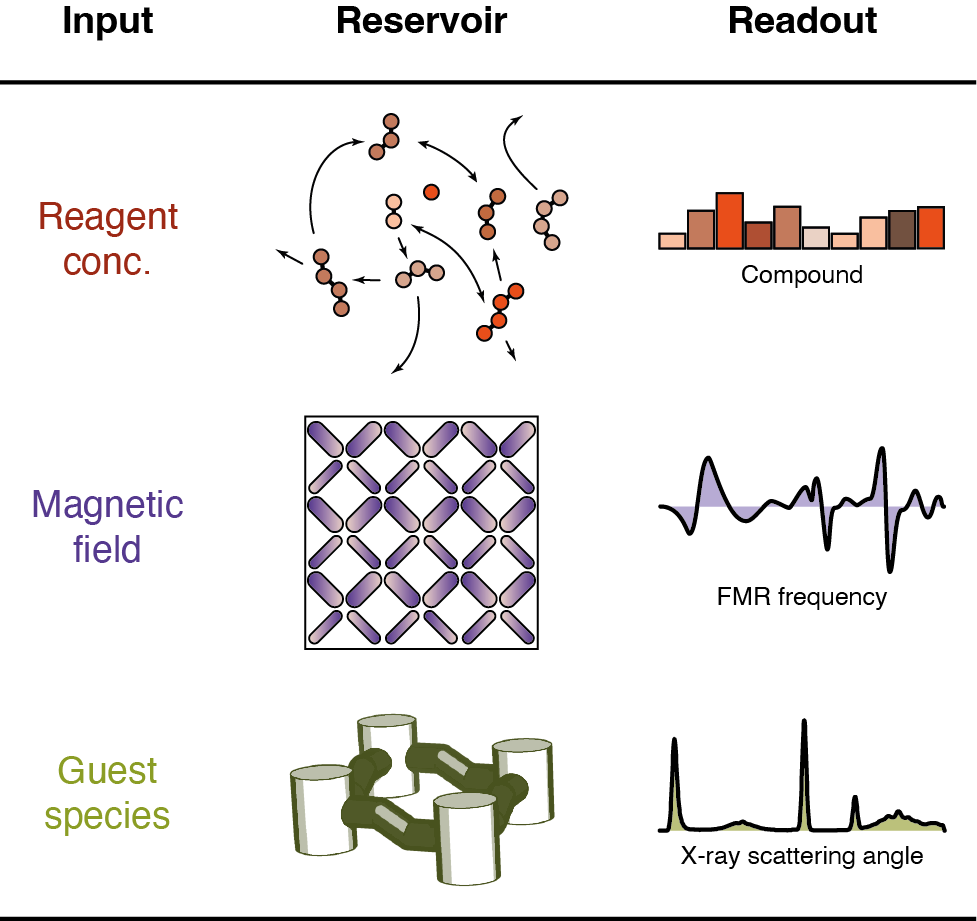}
	\caption{\footnotesize {\bf Examples of physical reservoirs.} In the formose chemical reservoir (top), a variety of polymeric species continuously interconvert. Varying the concentration of monomers or other reagents causes a shift in the composition of the mixture, which can be characterised by mass spectrometry.\cite{Baltussen_2024} By contrast, the artificial vortex spin ice reservoir (middle) is comprised of a nanoarray of magnets in varying magnetic states. Application of an external magnetic field shifts the system between  states, with a corresponding change in the ferromagnetic resonance (FMR) spectrum.\cite{Gartside_2022} Here we report the DUT-8 reservoir (bottom), where the positions of columnar pillars shift on adsorption of different guest species within framework pores. The configurational rearrangement leads to differences in the diffuse features of the X-ray diffraction pattern.}
	\label{fig1}
\end{figure}

The central challenge in that field is to control the nature and degree of complexity in physical reservoirs so as to optimise performance.\cite{Tanaka_2019} Here, two recent advances are particularly relevant. The first is the chemical reservoir developed in Ref.~\citenum{Baltussen_2024}, where the formose reaction is exploited to generate a dynamic library of small polymers from simple chemical ingredients. Variation in precursor concentrations behaves as input to which the system adjusts accordingly; mass spectrometry is then used to generate a readout of the system state at any given time [Fig.~\ref{fig1}]. Because the relationship between input and mixture composition is nonlinear, the system can solve nonlinear tasks such as classification problems with nontrivial decision boundaries. The second example is that of artificial spin-vortex ices,\cite{Gartside_2022} which are  metamaterials comprised of carefully-designed nanomagnet arrays. Here the input is magnetic field, which flips magnetisation states, while the readout is the spin-wave spectrum, itself sensitive to magnetic correlations [Fig.~\ref{fig1}]. There is again strong nonlinearity in response, and a fading memory effect, with the system capable of advanced machine learning tasks such as time-series transformations. These two examples demonstrate the prospect for reservoir computers based on chemical interactions---\emph{i.e.}\ information dense and open to chemical control---and that exploit specific geometric arrangements in solid-state devices.

This was the context that led us to question whether disordered MOFs might actually serve as effective physical reservoirs. Such systems offer atomic-scale complexity within a crystalline solid,\cite{Simonov_2020} and the chemical and geometric diversity of MOFs suggests potential control over reservoir characteristics.\cite{Yaghi_2003} Here we address this possibility by examining the specific case of DUT-8, chosen because it is the most prominent MOF for which disorder--disorder transitions have been demonstrated experimentally and are well characterised.\cite{Ehrling_2021} We begin our report with a precis of correlated disorder in DUT-8 and its response to external chemical stimulus. Our first key result is to show that X-ray diffuse scattering provides a suitable readout of DUT-8 system state, allowing application in reservoir computing. In particular, there is sufficiently strong nonlinearity that difficult classification tasks can be solved with accuracies comparable to those of established reservoir computing architectures. We then demonstrate that cycling of guest-driven structural changes in DUT-8 gives readouts with the requisite characteristics to allow nonlinear time-series transformations that depend explicitly on memory. As such, DUT-8 exhibits a `responsive disorder' with precisely the right characteristics for reservoir computing applications. We conclude by discussing the implications of our results both for the specific case of MOFs and also in the sense of exploiting responsive disorder in more general terms.

\section*{Responsive disorder in DUT-8}

\begin{figure}
    \centering
    \includegraphics{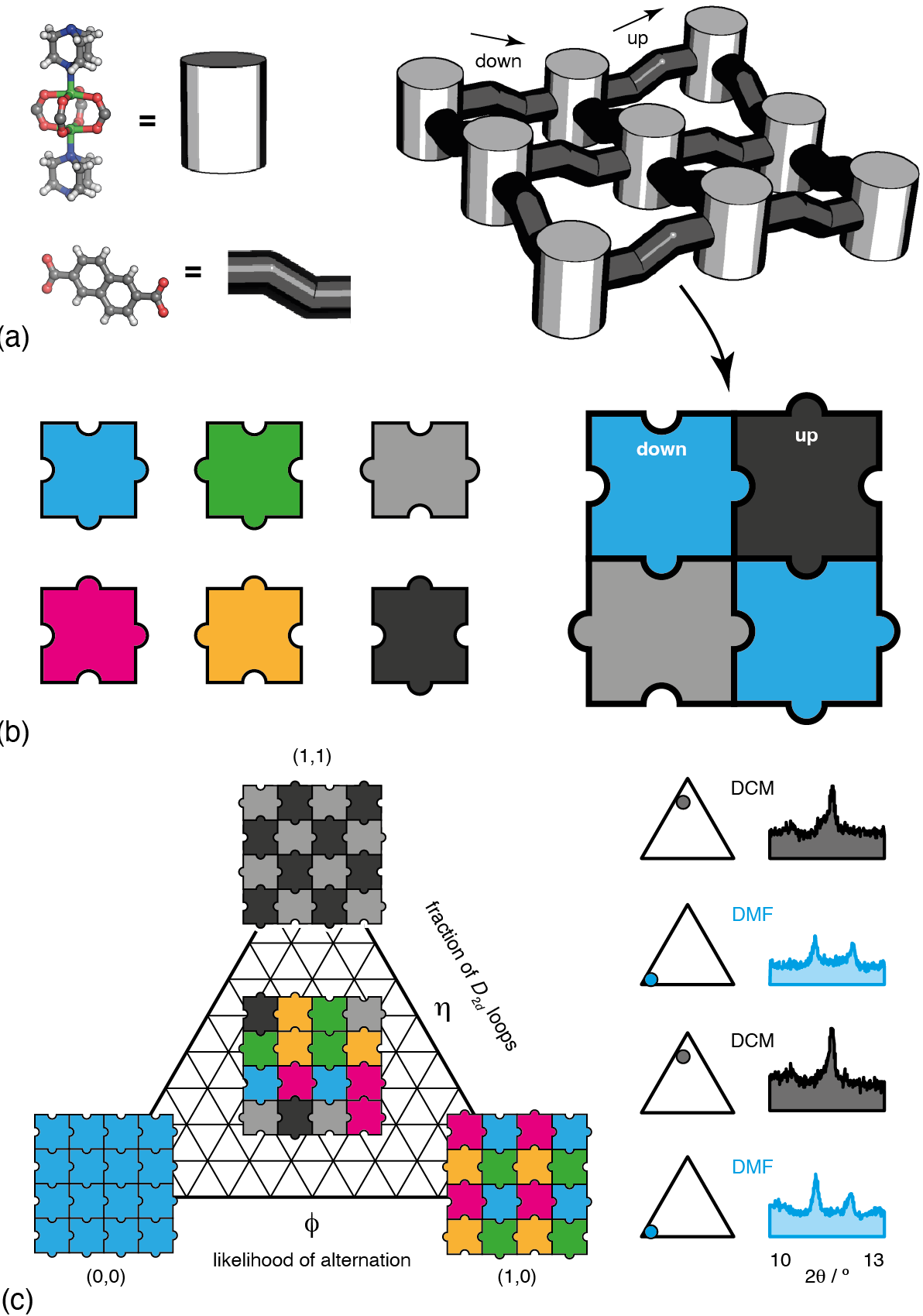}
	\caption{\footnotesize {\bf Correlated disorder in DUT-8.} (a) The framework structure of DUT-8 is assembled from dabco-linked nickel paddlewheels (shown as columns) connected by ndc linkers (shown as flexed tubes). Columns are arranged on a regular square lattice but at different relative heights: the ndc asymmetry shifts neighbouring columns. On traversing a square channel, one must encounter exactly two `up' shifts and two `down' shifts. (b) Jigsaw-tile representation of the six possible two-up-two-down channel arrangements, with $C_{2h}$ and $D_{2d}$ geometries shown in colour and grayscale, respectively. All DUT-8 configurations correspond to sensible jigsaw-tile arrangements. (c) Two-dimensional projection of the DUT-8 configurational landscape with local order parameters $(\phi,\eta)$. The boundary states are ordered; disordered arrangements are contained within the interior of the map. Interchanging the adsorbate in DUT-8 between DCM and DMF causes the system to traverse the configurational landscape, with a concomitant change in the X-ray diffraction pattern (data taken from Ref.~\citenum{Ehrling_2021}).}
	\label{fig2}
\end{figure}

DUT-8 is an ostensibly straightforward MOF assembled from columns of dabco-linked Ni$_2$-carboxylate paddlewheels.\cite{Klein_2010} These columns are arranged on a square grid, with neighbours connected via 2,6-naphthalene-dicarboxylate (ndc) linkers [Fig~\ref{fig2}(a)]. A peculiarity of the geometry of ndc is that each link necessarily induces a shift between neighbouring columns either parallel or antiparallel to the columnar axis.\cite{Petkov_2019,Reynolds_2021} The shift direction (`up' or `down', say) depends on ndc orientation. There is of course no energetic difference between the two shift directions for an isolated column pair,\cite{Petkov_2019} but as columns connect to form an extended framework, topological constraints of the extended DUT-8 architecture enforce strict collective rules that govern which particular shift combinations are possible in practice.\cite{Reynolds_2021} These rules can be understood by considering the trajectory of `up/down' shifts encountered on traversing a square channel of the DUT-8 structure: the natural chemical constraint that one must end up where one begins implies that every square-channel circuit includes exactly two `up' steps and exactly two `down' steps [Fig.~\ref{fig2}(a)]. Note that violation of this rule carries an enormous energy penalty in that it involves mismatch of not just a single carboxylate--paddlewheel connection but of entire columns of such connections.

There are exactly six channel configurations that satisfy the DUT-8 `two-up-two-down' constraint.\cite{Reynolds_2021,Ehrling_2021} These partition into one set of four corresponding to cyclic permutations of `up-up-down-down', and one set of two of `up-down-up-down' type. We use the visual language developed in Ref.~\citenum{Ehrling_2021} that represents these various cases by jigsaw tiles of different colours or grayscales [Fig.~\ref{fig2}(b)]. The key interpretation of this representation is that any sensible connecting arrangement of jigsaw tiles can be mapped onto a chemically feasible DUT-8 configuration. Limiting cases with ordered arrangements of jigsaw tiles correspond to DUT-8 states with well-defined crystal symmetries. These are the exception rather than the rule: there is a thermodynamically-dominant fraction of disordered configurations, with the (extensive) configurational entropy of the system as a whole given by Lieb's famous result of 1967.\cite{Lieb_1967b} In Fig.~\ref{fig2}(c), as in Ref.~\citenum{Ehrling_2021}, the diverse configurational landscape of possible DUT-8 states is projected onto a suitably compactified triangular map. It is this prior work that relates the phenomenology of DUT-8 to that of the so-called `six-vertex' model (jigsaw pieces map to vertex types), used historically to study magnetic and hydrogen-bond disorder in frustrated systems.\cite{Lieb_1967,Kondev_1997}

A key, and unexpected, result of Ref.~\citenum{Ehrling_2021} was the discovery that DUT-8 crystallites are not locked into a specific configuration during synthesis. Rather they can be converted reversibly from one configuration to another on exposure to different guest species. The chemical driving force for these disorder--disorder transformations is the interaction between guest species and pore geometries of different symmetries. As host--guest interactions vary, so too do the vertical positions of the paddlewheel--dabco columns and hence the longer-range DUT-8 structure. It is no surprise that the  experimental signature of each transition is clearest in the powder X-ray diffraction pattern,\cite{Keen_2015} where diffuse scattering signatures change as guest exchange induces structural reorganisation [Fig.~\ref{fig2}(c)]. Conceptually, this behaviour is crucial for reservoir computing because exposure to guest molecules can act as a tuneable `chemical field' that biases DUT-8 towards different regions of its configurational manifold---behaving as the chemical analogue of \emph{e.g.}\ magnetic field in vortex-ice reservoirs.\cite{Gartside_2022}

\section*{Results}

\begin{figure}
    \centering
    \includegraphics{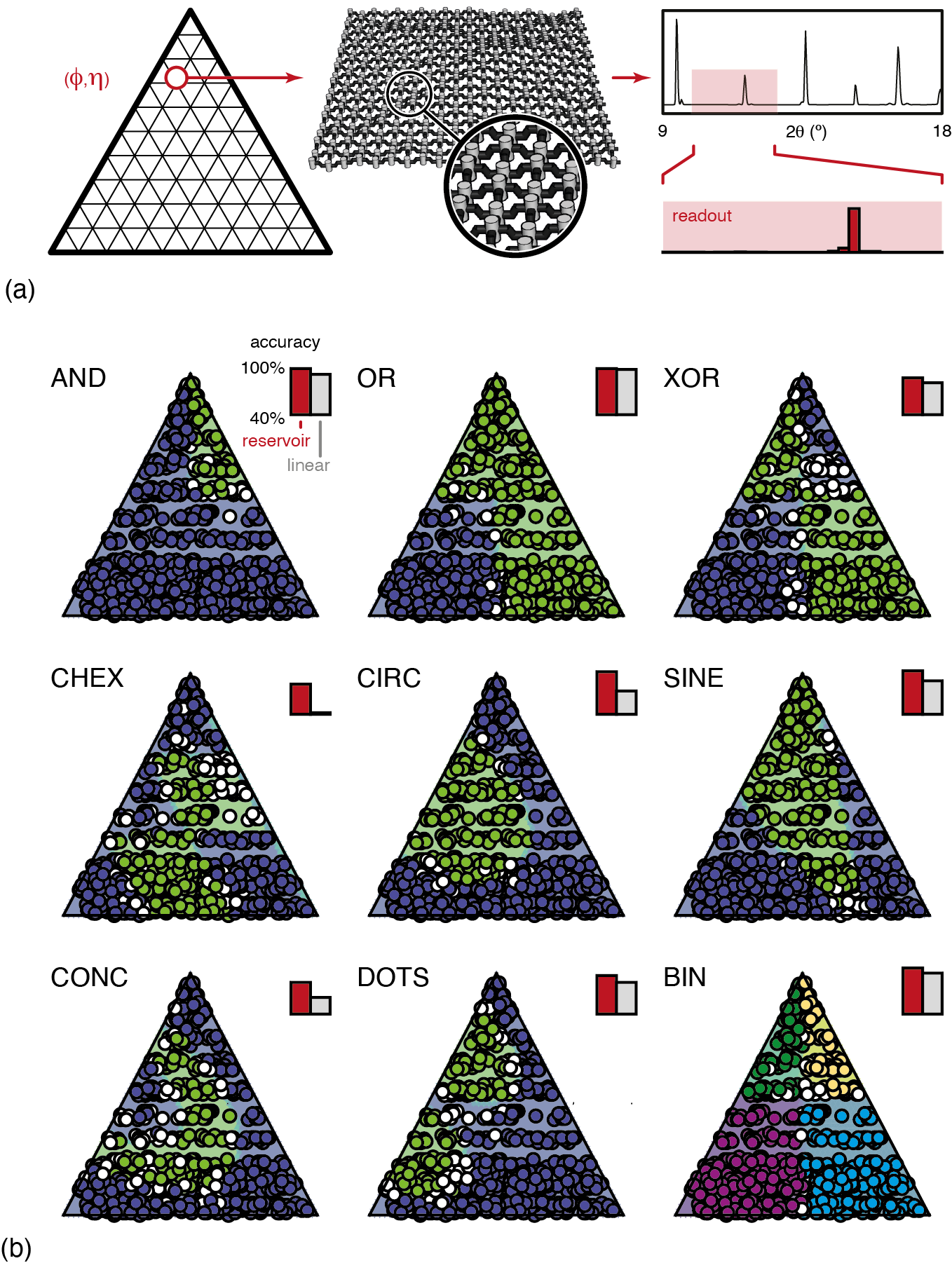}
	\caption{\footnotesize {\bf Classification tasks with a DUT-8 reservoir.} (a) Encoding of $(\phi,\eta)$ coordinates is carried out in three steps. First, a representative DUT-8 configuration is generated; second, the corresponding X-ray diffraction pattern is calculated; and third, a region of this diffraction pattern containing diffuse scattering is discretised to form a 27-channel readout. (b) Classification tasks involve using these readouts to train a single-layer linear support vector classifier against a range of target functions. Test results for each function are shown here, where coloured circles denote correct classification of different integral values. Incorrect classifications (white circles) tend to occur at the decision boundaries. Classification accuracy is always better than that for linear regression from input data (bar charts).}
	\label{fig3}
\end{figure}

Our starting point was to demonstrate that the mapping from DUT-8 configurational state to X-ray powder diffraction pattern is nonlinear in nature and that this nonlinearity is sufficient to be exploited usefully in machine learning tasks. We used two parameters to describe a given DUT-8 state: these were the $\phi$ and $\eta$ variables described initially in Ref.~\citenum{Ehrling_2021} and which act as coordinates for the 2D projection shown in Fig.~\ref{fig2}(c). Briefly, the parameter $0\leq\phi\leq1$ captures the tendency for up/down alternation in neighbouring square channels, and $0\leq\eta\leq1$ denotes the fraction of channels of the `up-down-up-down' type (\emph{i.e.} grayscale jigsaw tiles, $D_{2d}$ symmetry). There is an emergent geometric constraint that $\phi$ must be larger than $\eta$, but otherwise configurations with all states $(\phi,\eta)$ are realisable. We used a loop-move Monte Carlo (MC) algorithm\cite{Barkema_1998} to generate an ensemble of 3,000 configurations distributed across the DUT-8 landscape. Each configuration represented a $20\times20$ supercell of the parent DUT-8 structure and obeyed periodic boundary conditions. The choice of loop flips as move type is deliberate: they guarantee ergodicity,\cite{Barkema_1998} whilst ensuring that only physical plausible states are sampled. Each MC configuration was decorated with the coarse-grained DUT-8 representation of Ref.~\citenum{Ehrling_2021} and the corresponding X-ray diffraction pattern then calculated using TOPAS.\cite{Coelho_2016,Schmidt_2017} As system readout we used a small portion of these diffraction patterns corresponding to scattering angles $9.8\leq2\theta\leq 12.5^\circ$ ($\lambda=1.542$\,\AA) where the diffuse scattering dominates; the patterns were rebinned at $0.1^\circ$ intervals and normalised to obtain a 27-channel signature of each state [Fig.~\ref{fig3}(a)].

To test for nonlinearity we carried out a series of classification tasks. The MC configurations and corresponding diffuse-scattering patterns were first divided into train and test datasets. Each classification task then involved learning a well-defined response (\emph{e.g.}\ `0' or `1') that depended on the input location in $(\phi,\eta)$-space. A canonical example is the logical AND gate, where a single-layer classifier is trained to output 1 only when $\eta > \frac{1}{2}$ and $\phi>\frac{1}{2}(\eta+1)$ (\emph{i.e.}\ when the input lies in the top-right quadrant). For each classification task, we used the discretised diffraction patterns and target response for the training data set to train a linear support vector classifier that effectively mapped one onto the other. Note that this training step involved a single regression task that is both deterministic and computationally inexpensive.\cite{Tanaka_2019} The effectiveness of the resulting machine-learned model can be assessed by comparing calculated and ideal outputs for the test dataset. We show in Fig.~\ref{fig3}(b) the results for a series of standard classification tasks. The corresponding accuracies are remarkably good and are comparable to those obtained for the formose reservoir.\cite{Baltussen_2024} As is the case for all good reservoirs, errors usually only occur at the decision boundaries.\cite{Lukosevicius_2009}
 
But the clearest sign of nonlinearity comes from comparing the results of these classification tasks to those obtained without the DUT-8 reservoir. To do so one uses the $(\phi,\eta)$ coordinates themselves (\emph{i.e.}\ rather than the discretised diffraction patterns) together with the same target responses to train a new set of linear support vector classifiers. The resulting accuracies for the same train/test datasets used above then demonstrate the best possible outcomes for linear transformations; hence the marked improvement seen for the DUT-8 reservoir is attributable to the nonlinear high-dimensional mapping associated with its disordered states [Fig.~\ref{fig3}(b)]. This improvement is (understandably) greatest for the least linear classification tasks. From a chemical perspective, the fact that DUT-8 behaves in a nonlinear sense is entirely intuitive. As one changes the relative fractions of different channel types in its structure, one expects a complex variation in diffuse scattering: after all, the (Bragg) scattering patterns of the limiting ordered phases are not related by a linear transformation. Similar nontrivial variation in the form of diffuse scattering as a function of control parameter (\emph{e.g.}\ composition or temperature) has been widely observed in other materials classes, including Prussian blue analogues\cite{Simonov_2020b} and disordered rocksalts.\cite{Ji_2019}

Having established nonlinearity, we turned to characterising the memory characteristics of DUT-8. We used a variation of time-series transformations to do so, with our approach based heavily on that developed in Ref.~\citenum{Gartside_2022} to characterise spin-vortex ice reservoirs. In that work, the reservoir was subjected to a sinusoidally-varying magnetic field. As the system responded its spin-wave spectrum varied, and a single ridge regression layer could be trained to map these outputs onto a variety of different target functions.\cite{Gartside_2022} The existence of a memory component to reservoir response was demonstrated by successful learning of asymmetric functions where a common input field strength mapped onto different output values depending on the sequence of preceding states; the sawtooth function is a canonical example. Our DUT-8 implementation followed closely, with magnetic field being replaced by a solvent-driven `chemical field' and with discretised diffuse scattering patterns replacing the spin-wave spectra.

\begin{figure}
    \centering
    \includegraphics{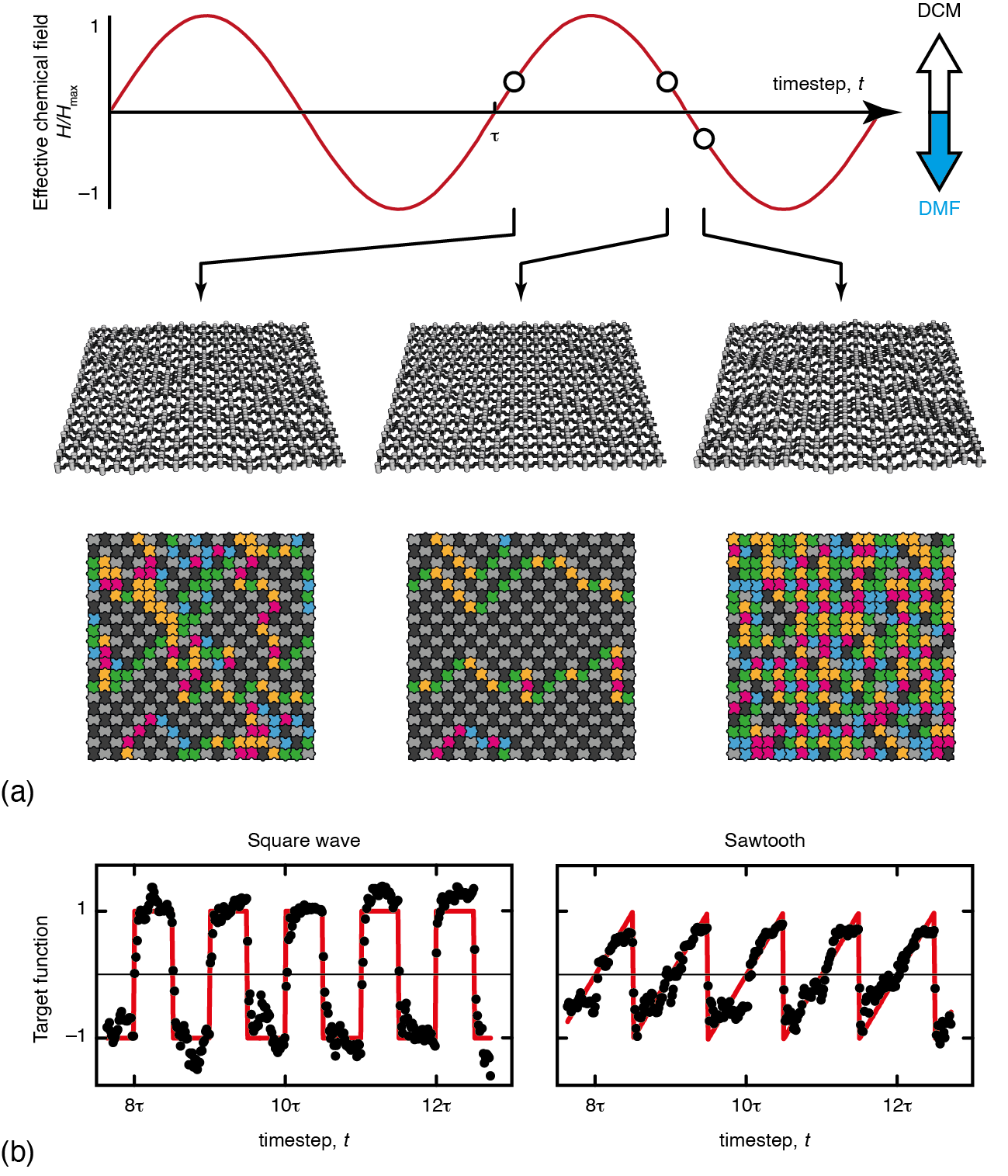}
	\caption{\footnotesize {\bf Time-series transformations using a DUT-8 reservoir.} To emulate the effect of changing adsorbate, the reservoir is subjected to a sinusoidally-varying effective chemical field. Monte Carlo simulations were carried out at each time step, driven by the Hamiltonian \eqref{hamil} and carrying forward configurations from step to step. Note that the DUT-8 state is history dependent: we illustrate this point with representative configurations (in both columnar and jigsaw representations) for two timesteps with equivalent applied fields (left and middle). On reversing the sign of the field, the system is driven towards low-$\eta$ configurations as observed experimentally on replacing DMF with DCM. (b) Representative symmetric (left) and asymmetric (right) time-series transformation test results. The target functions are shown as red lines and the machine-learned data as black circles.}
	\label{fig4}
\end{figure}

Our computational implementation involved the following steps. We included in our MC Hamiltonian a simple `chemical Zeeman' term
\begin{equation}
\mathcal H=-H\eta,\label{hamil}
\end{equation}
where $H$ is an effective field strength and $\eta$ is the coarse-grained measure of average pore geometry introduced above. This term is not intended as a microscopic description of host--guest energetics, but as a minimal effective interaction that captures the empirical preference of particular guests for specific pore geometries.  In this sense, $H$ acts as a chemical field conjugate to the disorder parameter $\eta$, directly analogous to the Zeeman effect in which magnetic field acts on magnetisation.\cite{Zeeman_1897} We modulated the applied field $H=H_{\rm max}\sin(2\pi t/\tau)$ to emulate guest exchange with time period $\tau$, carrying out the MC simulations for each time step $t$ at an effective temperature of $T_{\rm MC}=H_{\rm max}$ and for a constant number of proposed loop moves. This choice ensures that thermal fluctuations and field-driven bias are comparable in magnitude; we verified that the qualitative behaviour is insensitive to the precise simulation details (see Supporting Information). The resulting simulations behaved exactly as expected. So, for example, at any time $t$ for which $\sin(2\pi t/\tau)$ is at its maximum the MC simulations are heavily biased towards large-$\eta$ states, mimicking the experimental effect of incorporating DCM within the pores.\cite{Ehrling_2021} As $t$ varies thereafter and $H$ becomes negative, the bias reverses, favouring small-$\eta$ states, as observed experimentally when DCM is replaced by DMF [Fig.~\ref{fig4}(a)].\cite{Ehrling_2021}

We show in Fig.~\ref{fig4}(b) the results of two representative time-series transformations---one symmetric and one asymmetric---that were carried out using discretised diffuse scattering data from these MC simulations. To mirror as closely as possible the analysis of Ref.~\citenum{Gartside_2022}, we made use of 200 time steps for model training; details of the parameters involved, robustness testing, and discussion of the chaotic Mackey--Glass time series\cite{Mackey_1977} are given in the Supporting Information. Across nearly all examples we found similar performance to that of the vortex-spin ice reservoirs, with numerically comparable mean-squared deviations.\cite{Gartside_2022} In particular, the sawtooth function is faithfully reproduced, demonstrating the physically-encoded memory effect present within the DUT-8 system. That DUT-8 should have configurational memory is chemically reasonable, given that each guest-triggered reorganisation involves a collective shuffling of loops of dabco-linked columns, with the distribution of active loops dependent on system state.\cite{Oakes_2016} Experimentally, the data of Ref.~\citenum{Ehrling_2021} also show a subtle history-dependence in the diffuse scattering from DUT-8 during guest exchange cycles that supports this interpretation [Fig.~\ref{fig4}(a)]. We show in the Supporting Information that this memory fades with time, as observed (and exploited) in other physical reservoirs.\cite{Gartside_2022}

\section*{Discussion}

So why does DUT-8 behave so effectively as a physical reservoir? We argue that two ingredients are key: first, the presence of correlated disorder, and, second, the ability to manipulate that disorder in response to external stimuli. Transformations in ordered solid phases are often characterised by linearity (\emph{e.g.}\ Vegard's law, thermal expansion, compressibility)---both in terms of their structural response and the corresponding effect on diffraction patterns. At the opposite extreme, systems with uncorrelated disorder are typically well described by purely statistical models, and therefore lack the structured, history-dependent responses required for reservoir computing. Rather it is the complexity that emerges within the correlated disorder regime that endows DUT-8 with its interesting behaviour. Precisely these considerations have motivated the use of geometrically-frustrated artificial spin-ices as reservoirs by the mesoscale community.\cite{Ladak_2010} By analogy, we expect the behaviour demonstrated here to generalise to other solid phases that combine correlated disorder with responsiveness. We suggest the term `responsive disorder' to describe such scenarios,\cite{Goodwin_2025} with potential examples including liquid-crystal assemblies,\cite{Zeng_2011} hydrogen-bonded networks,\cite{MonterodeEspinosa_2017} and orbital-molecule liquids.\cite{Browne_2020,Griffiths_2024}

In the specific case of DUT-8, several immediate directions suggest themselves. The first is the use of \emph{in situ} X-ray diffraction measurements to determine the effectiveness of experimental scattering data in carrying out reservoir computing tasks.\cite{Bon_2015} Key practical considerations will include the kinetics of guest exchange, strain effects, and the role of particle size.\cite{Kavoosi_2017,Miura_2017} A second is to determine the link between MOF composition and computational performance, which might be achieved through \emph{e.g.}\ isoreticular chemistry.\cite{Ehrling_2019,Yaghi_2003} And a third is to explore the efficiencies of alternative inputs and readouts; \emph{e.g.}\ application of mechanical stress as input and infrared spectroscopy as output. Such approaches might suit other systems where configurational ordering is also strongly coupled to strain.\cite{Greenbaum_2024} Indeed there is no reason why other MOFs with responsive disorder shouldn't show even better performance in computing tasks---nor any reason why reservoir computing need be the only focus. More broadly, we hope that the results presented here illustrate how the complex configurational degeneracy accessible to correlated disordered states and the information content of diffuse scattering might be exploited in applications based on information processing and computing.

\section*{Acknowledgments}
A.L.G. thanks A. Anker (DTU),  R. I. Cooper (Oxford), L. Heydermann (ETH Zurich), and C. J. Kepert (Sydney) for useful discussions, and S. Kaskel (TU Dresden) for introducing him to the DUT-8 system. This research was supported financially by the European Research Council (Advanced Grant 788144), the UKRI through Frontier Research Grant EP/Z534031/1, and the Royal Society through the Faraday Discovery Fellowships Fund, provided by DSIT.

\section*{References}
\bibliography{arxiv_2026_reservoir}

\begin{thebibliography}{10}
\expandafter\ifx\csname url\endcsname\relax
  \def\url#1{\texttt{#1}}\fi
\expandafter\ifx\csname urlprefix\endcsname\relax\def\urlprefix{URL }\fi
\providecommand{\bibinfo}[2]{#2}
\providecommand{\eprint}[2][]{\url{#2}}

\bibitem{Weaver_1948}
\bibinfo{author}{Weaver, W.}
\newblock \bibinfo{title}{Science and complexity}.
\newblock \emph{\bibinfo{journal}{Am. Sci.}} \textbf{\bibinfo{volume}{36}},
  \bibinfo{pages}{536--544} (\bibinfo{year}{1948}).

\bibitem{Crutchfield_2011}
\bibinfo{author}{Crutchfield, J.~P.}
\newblock \bibinfo{title}{Between order and chaos}.
\newblock \emph{\bibinfo{journal}{Nat. Phys.}} \textbf{\bibinfo{volume}{8}},
  \bibinfo{pages}{17--24} (\bibinfo{year}{2011}).

\bibitem{Keen_2015}
\bibinfo{author}{Keen, D.~A.} \& \bibinfo{author}{Goodwin, A.~L.}
\newblock \bibinfo{title}{The crystallography of correlated disorder}.
\newblock \emph{\bibinfo{journal}{Nature}} \textbf{\bibinfo{volume}{521}},
  \bibinfo{pages}{303--309} (\bibinfo{year}{2015}).

\bibitem{Nield_1993}
\bibinfo{author}{Nield, V.~M.}, \bibinfo{author}{Keen, D.~A.},
  \bibinfo{author}{Hayes, W.} \& \bibinfo{author}{McGreevy, R.~L.}
\newblock \bibinfo{title}{Structure and fast-ion conduction in $\alpha$-{AgI}}.
\newblock \emph{\bibinfo{journal}{Solid State Ionics}}
  \textbf{\bibinfo{volume}{66}}, \bibinfo{pages}{247--258}
  (\bibinfo{year}{1993}).

\bibitem{Fennell_2009}
\bibinfo{author}{Fennell, T.} \emph{et~al.}
\newblock \bibinfo{title}{Magnetic coulomb phase in the spin ice
  {H}o$_2${T}i$_2${O}$_7$}.
\newblock \emph{\bibinfo{journal}{Science}} \textbf{\bibinfo{volume}{326}},
  \bibinfo{pages}{415--417} (\bibinfo{year}{2009}).

\bibitem{Blunt_2008}
\bibinfo{author}{Blunt, M.~O.} \emph{et~al.}
\newblock \bibinfo{title}{Random tiling and topological defects in a
  two-dimensional molecular network}.
\newblock \emph{\bibinfo{journal}{Science}} \textbf{\bibinfo{volume}{322}},
  \bibinfo{pages}{1077--1081} (\bibinfo{year}{2008}).

\bibitem{Meekel_2021}
\bibinfo{author}{Meekel, E.~G.} \& \bibinfo{author}{Goodwin, A.~L.}
\newblock \bibinfo{title}{Correlated disorder in metal--organic frameworks}.
\newblock \emph{\bibinfo{journal}{CrystEngComm}} \textbf{\bibinfo{volume}{23}},
  \bibinfo{pages}{2915--2922} (\bibinfo{year}{2021}).

\bibitem{Klein_2010}
\bibinfo{author}{Klein, N.} \emph{et~al.}
\newblock \bibinfo{title}{Monitoring adsorption-induced switching by
  $^{129}${X}e {NMR} spectroscopy in a new metal-organic framework
  {Ni$_2$(2,6-ndc)$_2$(dabco)}}.
\newblock \emph{\bibinfo{journal}{Phys. Chem. Chem. Phys.}}
  \textbf{\bibinfo{volume}{12}}, \bibinfo{pages}{11778--11784}
  (\bibinfo{year}{2010}).

\bibitem{Reynolds_2021}
\bibinfo{author}{Reynolds, E.~M.} \emph{et~al.}
\newblock \bibinfo{title}{Function from configurational degeneracy in
  disordered framework materials}.
\newblock \emph{\bibinfo{journal}{Faraday Discuss.}}
  \textbf{\bibinfo{volume}{225}}, \bibinfo{pages}{241--254}
  (\bibinfo{year}{2021}).

\bibitem{Lieb_1967}
\bibinfo{author}{Lieb, E.~H.}
\newblock \bibinfo{title}{Residual entropy of square ice}.
\newblock \emph{\bibinfo{journal}{Phys. Rev.}} \textbf{\bibinfo{volume}{162}},
  \bibinfo{pages}{162--172} (\bibinfo{year}{1967}).

\bibitem{Meekel_2023}
\bibinfo{author}{Meekel, E.~G.} \emph{et~al.}
\newblock \bibinfo{title}{Truchet-tile structure of a topologically aperiodic
  metal--organic framework}.
\newblock \emph{\bibinfo{journal}{Science}} \textbf{\bibinfo{volume}{379}},
  \bibinfo{pages}{357--361} (\bibinfo{year}{2023}).

\bibitem{Griffin_2025}
\bibinfo{author}{Griffin, S.~L.} \emph{et~al.}
\newblock \bibinfo{title}{A lanthanide {MOF} with nanostructured node
  disorder}.
\newblock \emph{\bibinfo{journal}{Nat. Commun.}} \textbf{\bibinfo{volume}{16}},
  \bibinfo{pages}{3209} (\bibinfo{year}{2025}).

\bibitem{Crutchfield_1994}
\bibinfo{author}{Crutchfield, J.~P.}
\newblock \bibinfo{title}{The calculi of emergence: computation, dynamics and
  induction}.
\newblock \emph{\bibinfo{journal}{Physica D}} \textbf{\bibinfo{volume}{75}},
  \bibinfo{pages}{11--54} (\bibinfo{year}{1994}).

\bibitem{Mandic_2001}
\bibinfo{editor}{Mandic, D.~P.} \& \bibinfo{editor}{Chambers, J.~A.} (eds.)
  \emph{\bibinfo{title}{Recurrent Neural Networks for Prediction: Learning
  Algorithms,Architectures and Stability}} (\bibinfo{publisher}{John Wiley \&
  Sons}, \bibinfo{address}{Chichester, UK}, \bibinfo{year}{2001}).

\bibitem{Jaeger_2004}
\bibinfo{author}{Jaeger, H.} \& \bibinfo{author}{Haas, H.}
\newblock \bibinfo{title}{Harnessing nonlinearity: Predicting chaotic systems
  and saving energy in wireless communication}.
\newblock \emph{\bibinfo{journal}{Science}} \textbf{\bibinfo{volume}{304}},
  \bibinfo{pages}{78--80} (\bibinfo{year}{2004}).

\bibitem{Lukosevicius_2009}
\bibinfo{author}{Luko{\v s}evi{\v c}ius, M.} \& \bibinfo{author}{Jaeger, H.}
\newblock \bibinfo{title}{Reservoir computing approaches to recurrent neural
  network training}.
\newblock \emph{\bibinfo{journal}{Comp. Sci. Rev.}}
  \textbf{\bibinfo{volume}{3}}, \bibinfo{pages}{127--149}
  (\bibinfo{year}{2009}).

\bibitem{Tanaka_2019}
\bibinfo{author}{Tanaka, G.} \emph{et~al.}
\newblock \bibinfo{title}{Recent advances in physical reservoir computing: {A}
  review}.
\newblock \emph{\bibinfo{journal}{Neural Networks}}
  \textbf{\bibinfo{volume}{115}}, \bibinfo{pages}{100--123}
  (\bibinfo{year}{2019}).

\bibitem{Soriano_2015}
\bibinfo{author}{Soriano, M.~C.} \emph{et~al.}
\newblock \bibinfo{title}{Delay-based reservoir computing: Noise effects in a
  combined analog and digital implementation}.
\newblock \emph{\bibinfo{journal}{IEEE Trans. Neur. Net. Learn. Sys.}}
  \textbf{\bibinfo{volume}{26}}, \bibinfo{pages}{388--393}
  (\bibinfo{year}{2015}).

\bibitem{Caluwaerts_2014}
\bibinfo{author}{Caluwaerets, K.} \emph{et~al.}
\newblock \bibinfo{title}{Design and control of compliant tensegrity robots
  through simulation and hardware validation}.
\newblock \emph{\bibinfo{journal}{J. R. Soc. Interface}}
  \textbf{\bibinfo{volume}{11}}, \bibinfo{pages}{2014520}
  (\bibinfo{year}{2014}).

\bibitem{Nakajima_2015}
\bibinfo{author}{Nakajima, K.}, \bibinfo{author}{Hauser, H.},
  \bibinfo{author}{Li, T.} \& \bibinfo{author}{Pfeifer, R.}
\newblock \bibinfo{title}{Information processing via physical soft body}.
\newblock \emph{\bibinfo{journal}{Sci. Rep.}} \textbf{\bibinfo{volume}{5}},
  \bibinfo{pages}{10487} (\bibinfo{year}{2015}).

\bibitem{Baltussen_2024}
\bibinfo{author}{Baltussen, M.~G.}, \bibinfo{author}{de~Jong, T.~J.},
  \bibinfo{author}{Duez, Q.}, \bibinfo{author}{Robinson, W.~E.} \&
  \bibinfo{author}{Huck, W. T.~S.}
\newblock \bibinfo{title}{Chemical reservoir computation in a self-organizing
  reaction network}.
\newblock \emph{\bibinfo{journal}{Nature}} \textbf{\bibinfo{volume}{631}},
  \bibinfo{pages}{549--555} (\bibinfo{year}{2024}).

\bibitem{Gartside_2022}
\bibinfo{author}{Gartside, J.~C.} \emph{et~al.}
\newblock \bibinfo{title}{Reconfigurable training and reservoir computing in an
  artificial spin-vortex ice via spin-wave fingerprinting}.
\newblock \emph{\bibinfo{journal}{Nat. Nano.}} \textbf{\bibinfo{volume}{17}},
  \bibinfo{pages}{460--469} (\bibinfo{year}{2022}).

\bibitem{Simonov_2020}
\bibinfo{author}{Simonov, A.} \& \bibinfo{author}{Goodwin, A.~L.}
\newblock \bibinfo{title}{Designing disorder into crystalline materials}.
\newblock \emph{\bibinfo{journal}{Nat. Rev. Chem.}}
  \textbf{\bibinfo{volume}{4}}, \bibinfo{pages}{657--673}
  (\bibinfo{year}{2020}).

\bibitem{Yaghi_2003}
\bibinfo{author}{Yaghi, O.~M.} \emph{et~al.}
\newblock \bibinfo{title}{Reticular synthesis and the design of new materials}.
\newblock \emph{\bibinfo{journal}{Nature}} \textbf{\bibinfo{volume}{423}},
  \bibinfo{pages}{705--714} (\bibinfo{year}{2003}).

\bibitem{Ehrling_2021}
\bibinfo{author}{Ehrling, S.} \emph{et~al.}
\newblock \bibinfo{title}{Adaptive response of a metal--organic framework
  through reversible disorder--disorder transitions}.
\newblock \emph{\bibinfo{journal}{Nat. Chem.}} \textbf{\bibinfo{volume}{13}},
  \bibinfo{pages}{568--574} (\bibinfo{year}{2021}).

\bibitem{Petkov_2019}
\bibinfo{author}{Petkov, P.} \emph{et~al.}
\newblock \bibinfo{title}{Conformational isomerism controls collective
  flexibility in metal-organic framework {DUT-8(Ni)}}.
\newblock \emph{\bibinfo{journal}{Phys. Chem. Chem. Phys.}}
  \textbf{\bibinfo{volume}{21}}, \bibinfo{pages}{674--680}
  (\bibinfo{year}{2019}).

\bibitem{Lieb_1967b}
\bibinfo{author}{Lieb, E.~H.}
\newblock \bibinfo{title}{Exact solution of the problem of the entropy of
  two-dimensional ice}.
\newblock \emph{\bibinfo{journal}{Phys. Rev. Lett.}}
  \textbf{\bibinfo{volume}{18}}, \bibinfo{pages}{692--694}
  (\bibinfo{year}{1967}).

\bibitem{Kondev_1997}
\bibinfo{author}{Kondev, J.}
\newblock \bibinfo{title}{Liouville field theory of fluctuating loops}.
\newblock \emph{\bibinfo{journal}{Phys. Rev. Lett.}}
  \textbf{\bibinfo{volume}{78}}, \bibinfo{pages}{4320--4323}
  (\bibinfo{year}{1997}).

\bibitem{Barkema_1998}
\bibinfo{author}{Barkema, G.~T.} \& \bibinfo{author}{Newman, M. E.~J.}
\newblock \bibinfo{title}{{Monte Carlo} simulation of ice models}.
\newblock \emph{\bibinfo{journal}{Phys. Rev. E}} \textbf{\bibinfo{volume}{57}},
  \bibinfo{pages}{1155--1166} (\bibinfo{year}{1998}).

\bibitem{Coelho_2016}
\bibinfo{author}{Coelho, A.~A.}
\newblock \bibinfo{title}{{TOPAS-Academic}, version 6} \bibinfo{pages}{Coelho
  Software: Brisbane} (\bibinfo{year}{2016}).

\bibitem{Schmidt_2017}
\bibinfo{author}{Schmidt, E.} \& \bibinfo{author}{Neder, R.~B.}
\newblock \bibinfo{title}{Diffuse single-crystal scattering corrected for
  molecular form factor effects}.
\newblock \emph{\bibinfo{journal}{Acta Cryst. A}}
  \textbf{\bibinfo{volume}{73}}, \bibinfo{pages}{231--237}
  (\bibinfo{year}{2017}).

\bibitem{Simonov_2020b}
\bibinfo{author}{Simonov, A.} \emph{et~al.}
\newblock \bibinfo{title}{Hidden diversity of vacancy networks in {P}russian
  blue analogues}.
\newblock \emph{\bibinfo{journal}{Nature}} \textbf{\bibinfo{volume}{578}},
  \bibinfo{pages}{256--260} (\bibinfo{year}{2020}).

\bibitem{Ji_2019}
\bibinfo{author}{Ji, H.} \emph{et~al.}
\newblock \bibinfo{title}{Hidden structural and chemical order controls lithium
  transport in cation-disordered oxides for rechargeable batteries}.
\newblock \emph{\bibinfo{journal}{Nat. Commun.}} \textbf{\bibinfo{volume}{10}},
  \bibinfo{pages}{592} (\bibinfo{year}{2019}).

\bibitem{Zeeman_1897}
\bibinfo{author}{Zeeman, P.}
\newblock \bibinfo{title}{The effect of magnetisation on the nature of light
  emitted by a substance}.
\newblock \emph{\bibinfo{journal}{Nature}} \textbf{\bibinfo{volume}{55}},
  \bibinfo{pages}{347} (\bibinfo{year}{1897}).

\bibitem{Mackey_1977}
\bibinfo{author}{Mackey, M.~C.} \& \bibinfo{author}{Glass, L.}
\newblock \bibinfo{title}{Oscillation and chaos in physiological control
  systems}.
\newblock \emph{\bibinfo{journal}{Science}} \textbf{\bibinfo{volume}{197}},
  \bibinfo{pages}{287--289} (\bibinfo{year}{1977}).

\bibitem{Oakes_2016}
\bibinfo{author}{Oakes, T.}, \bibinfo{author}{Garrahan, J.~P.} \&
  \bibinfo{author}{Powell, S.}
\newblock \bibinfo{title}{Emergence of cooperative dynamics in fully packed
  classical dimers}.
\newblock \emph{\bibinfo{journal}{Phys. Rev. E}} \textbf{\bibinfo{volume}{93}},
  \bibinfo{pages}{032129} (\bibinfo{year}{2016}).

\bibitem{Ladak_2010}
\bibinfo{author}{Ladak, S.}, \bibinfo{author}{Read, D.~E.},
  \bibinfo{author}{Perkins, G.~K.}, \bibinfo{author}{Cohen, L.~F.} \&
  \bibinfo{author}{Branford, W.~R.}
\newblock \bibinfo{title}{Direct observation of magnetic monopole defects in an
  artificial spin-ice system}.
\newblock \emph{\bibinfo{journal}{Nat. Phys.}} \textbf{\bibinfo{volume}{6}},
  \bibinfo{pages}{359--363} (\bibinfo{year}{2010}).

\bibitem{Goodwin_2025}
\bibinfo{author}{Goodwin, A.~L.}
\newblock \bibinfo{title}{Structural complexity and correlated disorder in
  materials chemistry}.
\newblock \emph{\bibinfo{journal}{arXiv:}} \bibinfo{pages}{2509.09171}
  (\bibinfo{year}{2025}).

\bibitem{Zeng_2011}
\bibinfo{author}{Zeng, X.} \emph{et~al.}
\newblock \bibinfo{title}{Complex multicolor tilings and critical phenomena in
  tetraphilic liquid crystals}.
\newblock \emph{\bibinfo{journal}{Science}} \textbf{\bibinfo{volume}{331}},
  \bibinfo{pages}{1302--1306} (\bibinfo{year}{2011}).

\bibitem{MonterodeEspinosa_2017}
\bibinfo{author}{Montero~de Espinosa, L.}, \bibinfo{author}{Meesorn, W.},
  \bibinfo{author}{Moatsou, D.} \& \bibinfo{author}{Weder, C.}
\newblock \bibinfo{title}{Bioinspired polymer systems with stimuli-responsive
  mechanical properties}.
\newblock \emph{\bibinfo{journal}{Chem. Rev.}} \textbf{\bibinfo{volume}{117}}
  (\bibinfo{year}{2017}).

\bibitem{Browne_2020}
\bibinfo{author}{Browne, A.~J.} \& \bibinfo{author}{Attfield, J.~P.}
\newblock \bibinfo{title}{Orbital molecules in vanadium oxide spinels}.
\newblock \emph{\bibinfo{journal}{Phys. Rev. B}}
  \textbf{\bibinfo{volume}{101}}, \bibinfo{pages}{024112}
  (\bibinfo{year}{2020}).

\bibitem{Griffiths_2024}
\bibinfo{author}{Griffiths, J.} \emph{et~al.}
\newblock \bibinfo{title}{Resolving length-scale-dependent transient disorder
  through an ultrafast phase transition}.
\newblock \emph{\bibinfo{journal}{Nat. Mater.}} \textbf{\bibinfo{volume}{23}},
  \bibinfo{pages}{1041--1047} (\bibinfo{year}{2024}).

\bibitem{Bon_2015}
\bibinfo{author}{Bon, V.} \emph{et~al.}
\newblock \bibinfo{title}{Exceptional adsorption-induced cluster and network
  deformation in the flexible metal-organic framework {DUT-8(Ni)} observed by
  {\it in situ} {X}-ray diffraction and {EXAFS}}.
\newblock \emph{\bibinfo{journal}{Phys. Chem. Chem. Phys.}}
  \textbf{\bibinfo{volume}{17}}, \bibinfo{pages}{17471--17449}
  (\bibinfo{year}{2015}).

\bibitem{Kavoosi_2017}
\bibinfo{author}{Kavoosi, N.} \emph{et~al.}
\newblock \bibinfo{title}{Tailoring adsorption induced phase transitions in the
  pillared-layer type metal-organic framework {DUT-8(Ni)}}.
\newblock \emph{\bibinfo{journal}{Dalton Trans.}}
  \textbf{\bibinfo{volume}{46}}, \bibinfo{pages}{4685--4695}
  (\bibinfo{year}{2017}).

\bibitem{Miura_2017}
\bibinfo{author}{Miura, H.} \emph{et~al.}
\newblock \bibinfo{title}{Tuning the gate-opening pressure and particle size
  distribution of the switchable metal-organic framework {DUT-8(Ni)} by
  controlled nucleation in a micromixer}.
\newblock \emph{\bibinfo{journal}{Dalton Trans.}}
  \textbf{\bibinfo{volume}{46}}, \bibinfo{pages}{14002--14011}
  (\bibinfo{year}{2017}).

\bibitem{Ehrling_2019}
\bibinfo{author}{Ehrling, S.} \emph{et~al.}
\newblock \bibinfo{title}{Crystal size {\it versus} paddle wheel deformbility:
  selective gated adsorption transitions of the switchable metal-organic
  frameworks {DUT-8(Co)} and {DUT-8(Ni)}}.
\newblock \emph{\bibinfo{journal}{J. Mater. Chem. A}}
  \textbf{\bibinfo{volume}{7}}, \bibinfo{pages}{21459--21475}
  (\bibinfo{year}{2019}).

\bibitem{Greenbaum_2024}
\bibinfo{author}{Greenbaum, G.} \emph{et~al.}
\newblock \bibinfo{title}{{\it In situ} observation of topotactic linker
  reorganization in the aperiodic metal--organic framework {TRUMOF-1}}.
\newblock \emph{\bibinfo{journal}{J. Am. Chem. Soc.}}
  \textbf{\bibinfo{volume}{146}}, \bibinfo{pages}{27262--27266}
  (\bibinfo{year}{2024}).

\end{thebibliography}

\end{document}